\documentclass[aps,prb,twocolumn,showpacs,superscriptaddress]{revtex4}

\usepackage{graphicx}

\begin{document}

\title{Elucidating the magnetic and superconducting phases in the alkali metal intercalated iron chalcogenides}
\author{Meng Wang}
\email{wangm@berkeley.edu}
\affiliation{Department of Physics, University of California, Berkeley, CA 94720, USA }
\author{Ming Yi}
\affiliation{Department of Physics, University of California, Berkeley, CA 94720, USA }
\author{Wei Tian}
\affiliation{Quantum Condensend Matter Division, Oak Ridge National Laboratory, Oak Ridge, TN 37831, USA}
\author{Edith Bourret-Courchesne}
\affiliation{Materials Science Division, Lawrence Berkeley National Laboratory, Berkeley, CA 94720, USA }
\author{Robert J. Birgeneau}
\affiliation{Department of Physics, University of California, Berkeley, CA 94720, USA }
\affiliation{Materials Science Division, Lawrence Berkeley National Laboratory, Berkeley, CA 94720, USA }
\affiliation{Department of Materials Science and Engineering, University of California, Berkeley, CA 94720, USA }

\begin{abstract}
The complex interdigitated phases have greatly frustrated attempts to document the basic features of the superconductivity in the alkali metal intercalated iron chalcogenides.  Here, using elastic neutron scattering, energy-dispersive x-ray spectroscopy, and resistivity measurements, we elucidate the relations of these phases in Rb$_{1-\delta}$Fe$_y$Se$_{2-z}$S$_z$. We find: i) the iron content is crucial in stabilizing the stripe antiferromagnetic (AF) phase with rhombic iron vacancy order ($y\approx1.5$), the block AF phase with $\sqrt{5}\times \sqrt{5}$ iron vacancy order ($y\approx1.6$), and the iron vacancy-free phase ($y\approx2$); ii) the superconducting phase ($z=0$) evolves into a metallic phase ($z>1.5$) with sulfur substitution due to the progressive decrease of the electronic correlation strength. Both the stripe AF phase and the block AF phase are Mott insulators. Our data suggest that there are miscibility gaps between these three phases. The existence of the miscibility gaps in the iron content is the key to understanding the relationship between these complicated phases.
\end{abstract}

\maketitle



\section{Introduction}
\begin{figure}[b]
\includegraphics[scale=0.4]{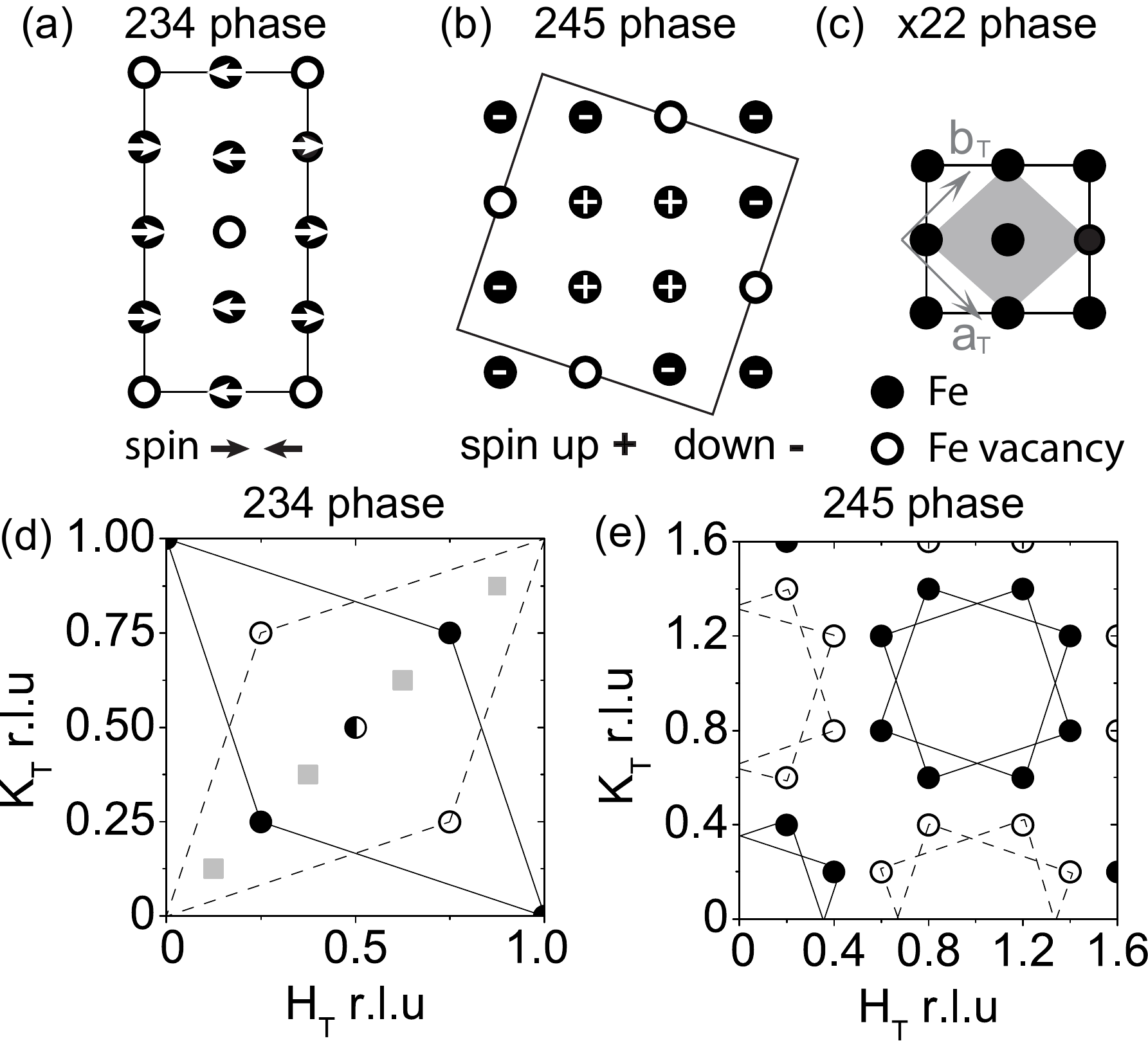}
\caption{(a) Schematics of an iron plane with in-plane ordered spins and rhombic iron vacancy order of the 234 phase, (b) the $c$ - axis ordered spins and $\sqrt{5} \times \sqrt{5}$ iron vacancy order of the 245 phase, and (c) the iron vacancy-free ``$\mathrm{x22}$" phase. The shaded area in (c) represents the tetragonal unit cell we used. (d) The filled and open circles mark the wave vectors of reflection peaks associated with the 234 phase and (e) the 245 phase in the reciprocal space corresponding to the tetragonal unit cell that is used throughout the paper.  The light gray squares in (d) show the wave vectors of a new set of peaks observed in Rb$_{0.8}$Fe$_{1.5}$Se$_2$.  }
\label{fig0}
\end{figure}

The discovery of superconductivity in the iron chalcogenides has generated a great deal of interest.  The simplest PbO-type bulk FeSe superconducts below 8 K and 36.7 K under ambient and 8.9 GPa pressure, respectively, whereas single-layer FeSe films on a SrTiO$_3$ substrate have been reported to have superconducting (SC) transition temperatures ($T_c$) as high as 109 K\cite{Hsu2008, Medvedev2009,Ge2014}. Intercalation of alkali-metals ($A =$ Li, Na, K, Rb, and Cs) or molecular spacer layer could enhance the $T_c$ in bulk materials up to 46 K\cite{Guo2010,Burrard2012,Ying2012,Lu2014b}. Most of the iron-based superconductors in bulk materials have been shown to be in the vicinity of an antiferromagnetically ordered parent compound. Furthermore, it has been found that the AF parent compounds could be progressively tuned to superconductors via carrier doping, isovalent doping, or pressure\cite{Johnston2010,Stewart2011}. Consequently, the AF ground state is viewed as one universal characteristic of the parent compounds in the iron pnictide and cuprate superconductors. In the alkali-metal intercalated iron chalcogenide SC system, $A_{1-\delta}$Fe$_{y}$Se$_2$, a block AF order with large moments of 3.3$\mu_B$ and $\sqrt{5}\times \sqrt{5}$ iron vacancy order as presented in Fig.~\ref{fig0} ($y\approx1.6$, referred to as the 245 phase) always seems to be mesoscopically interdigitated with the SC phase in the studies to-date\cite{Bao2011,Wang2012a,Ricci2011,Charnukha2012,Chen2011a,Ksenofontov2011,Texier2012,Gao2014,Ye2014,Wang2015a,Li2012}. In addition, a stripe AF order that has the same magnetic structure as that in the iron pnictide parent materials albeit with additional rhombic iron vacancy order ($y\approx1.5$, referred to as the 234 phase) and a phase with one Fe vacancy in eight ($y\approx1.75$, referred to as the  278 phase) have also been proposed as possible parent compounds\cite{Zhao2012,Ding2013}. However, each of the iron content, carrier doping, and isovalent substitution affects the formation of the various phases and the $T_c$s\cite{Gu2012,Wang2013a,Yan2012,Dagotto2013}. Thus, it is crucial to establish phase diagrams as a function of parameters that affect the superconductivity and compare them with those of the iron pnictide and cuprate superconductors in order to identify the nature of the magnetic and SC phases and to elucidate the relationship between them.

\begin{figure*}[t]
\includegraphics[scale=0.63]{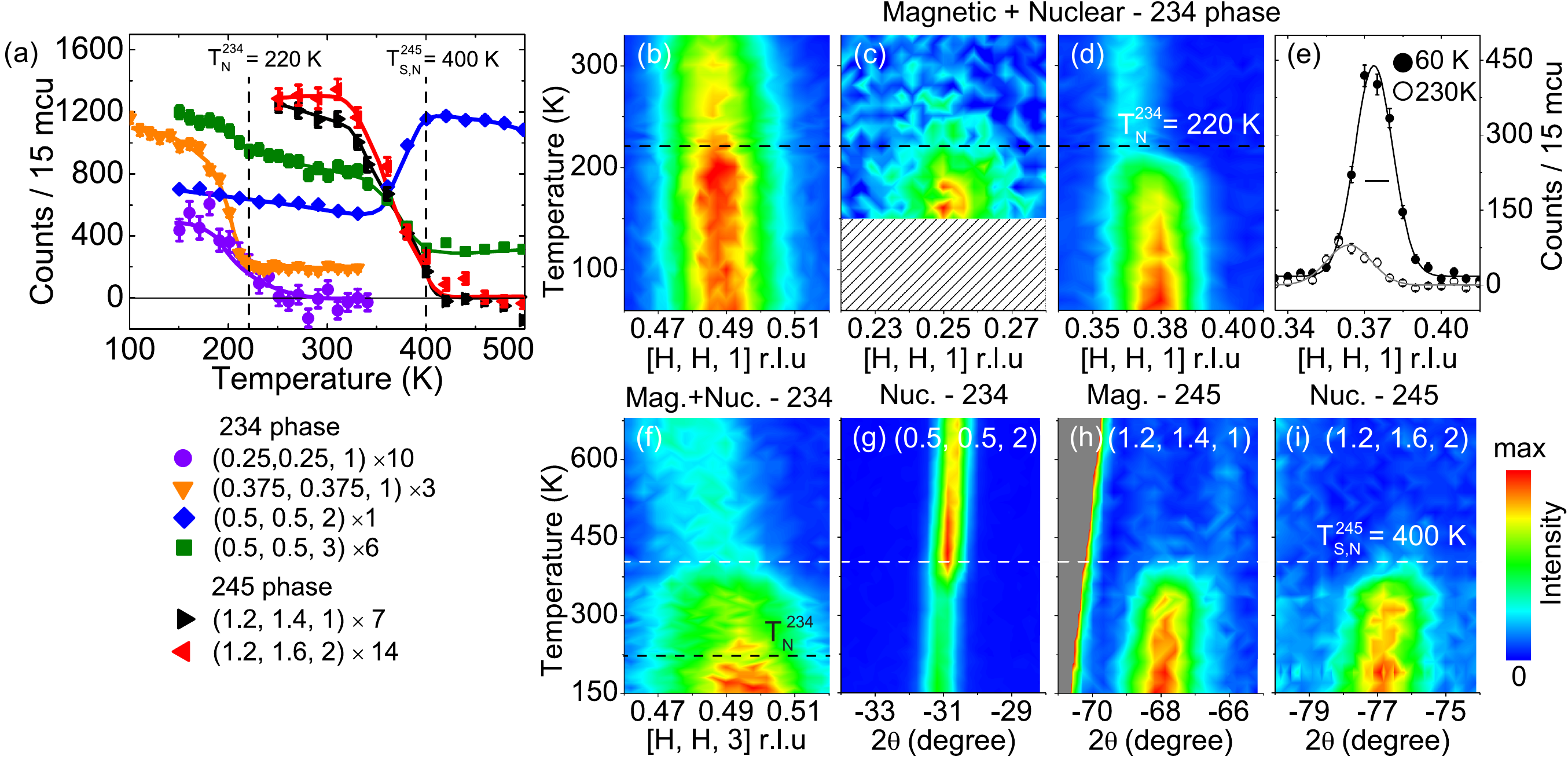}
\caption{(Color online). (a) Temperature dependence of the peak intensities in Rb$_{0.8}$Fe$_{1.5}$Se$_2$. The intensities for each peak have been normalized by a factor as shown following the corresponding wave vector. (b)-(f) The magnetic and superlattice peaks of the 234 phase scanned along the $[H, H, 1]$ direction and (f) the $[H, H, 3]$ direction. The bar in (e) indicates the instrumental resolution. (g) $\theta-2\theta$ scans of the rhombic iron vacancy order for the 234 phase at Q = (0.5, 0.5, 2) as a function of temperature. (h) $\theta-2\theta$ scans of the 245 phase through the magnetic peak $Q=(1.2, 1.4, 1)$ and (i) the $\sqrt{5} \times\sqrt{5}$ iron vacancy ordered peak $Q=(1.2, 1.6, 2)$. The black and white dashed lines mark the N$\acute{\mathrm{e}}$el temperature of the 234 phase at $T_N^{234}=220$ K and the coincidental magnetic and structure transition temperature of the 245 phase at  $T_{s,N}^{245}=400$ K, respectively. The intensities in (b-i) have been normalized for better comparison.  The lattice constants are chosen as the average of the two phases with $a = b = 3.991$ \AA\, and $c = 14.017$ \AA\ in the tetragonal notation optimized at 60 K.  }

\label{fig1}
\end{figure*}

For these purposes, we have synthesized different compositions of Rb$_{1-\delta}$Fe$_y$Se$_{2-z}$S$_z$ single crystals to study the effects of varying the Fe content and the Se:S ratio on the magnetic and superconducting phases. Firstly, using elastic neutron scattering, we find that the 234 phase and the 245 phase exist in all three insulating nominal compositions of Rb$_{0.8}$Fe$_{1.5}$Se$_2$, Rb$_{0.8}$Fe$_{1.5}$SeS, and Rb$_{0.8}$Fe$_{1.5}$S$_2$\cite{Wangm2014}, demonstrating that it is the Fe content rather than the Se:S ratio that is crucial to the formation of these phases. Furthermore, varying the iron content in the range of $1.5\leq y \leq1.6$ is found to affect only the relative volume fractions of the two phases rather than the characteristics of each structure. This necessitates that there is a miscibility gap between the two phases existing at $y\approx1.5$ and 1.6. 

The nominal composition of Rb$_{0.8}$Fe$_2$S$_2$ exhibits a single phase with randomly distributed iron vacancies at high temperatures. At temperatures below $T_s = 554$ K, it separates into two phases in a first order transition: the insulating 245 phase with $\sqrt{5}\times \sqrt{5}$ iron vacancy order and a metallic iron vacancy-free phase, which evolves into the SC phase in Rb$_{0.8}$Fe$_2$Se$_2$ with replacement of S by Se. The first order structural transition and the significant difference of the bond lengths of the two phases is consistent with the existence of a miscibility gap between the 245 phase and a metallic phase at $y\approx1.6$ and 2, respectively. The resulting phase diagrams show that the iron content is a key parameter in tuning between and stabilizing the insulating and metallic phases, while the Se:S ratio is important for inducing superconductivity from the non-magnetic metallic phase of Rb$_{1-\delta}$Fe$_2$S$_{2}$.

\section{Experiments}

We carried out the neutron scattering experiments on the HB-1A triple axis spectrometer at the High-Flux Isotope Reactor, Oak Ridge National Laboratory. The details about the single crystal growth, resistivity measurements, and configurations of neutron scattering experiments have been described elsewhere\cite{Wangm2014}. The electron beam for the energy-dispersive x-ray spectroscopy (EDX) was fixed at 20.0 eV.

\section{Results}

\begin{figure*}[t]
\includegraphics[scale=0.65]{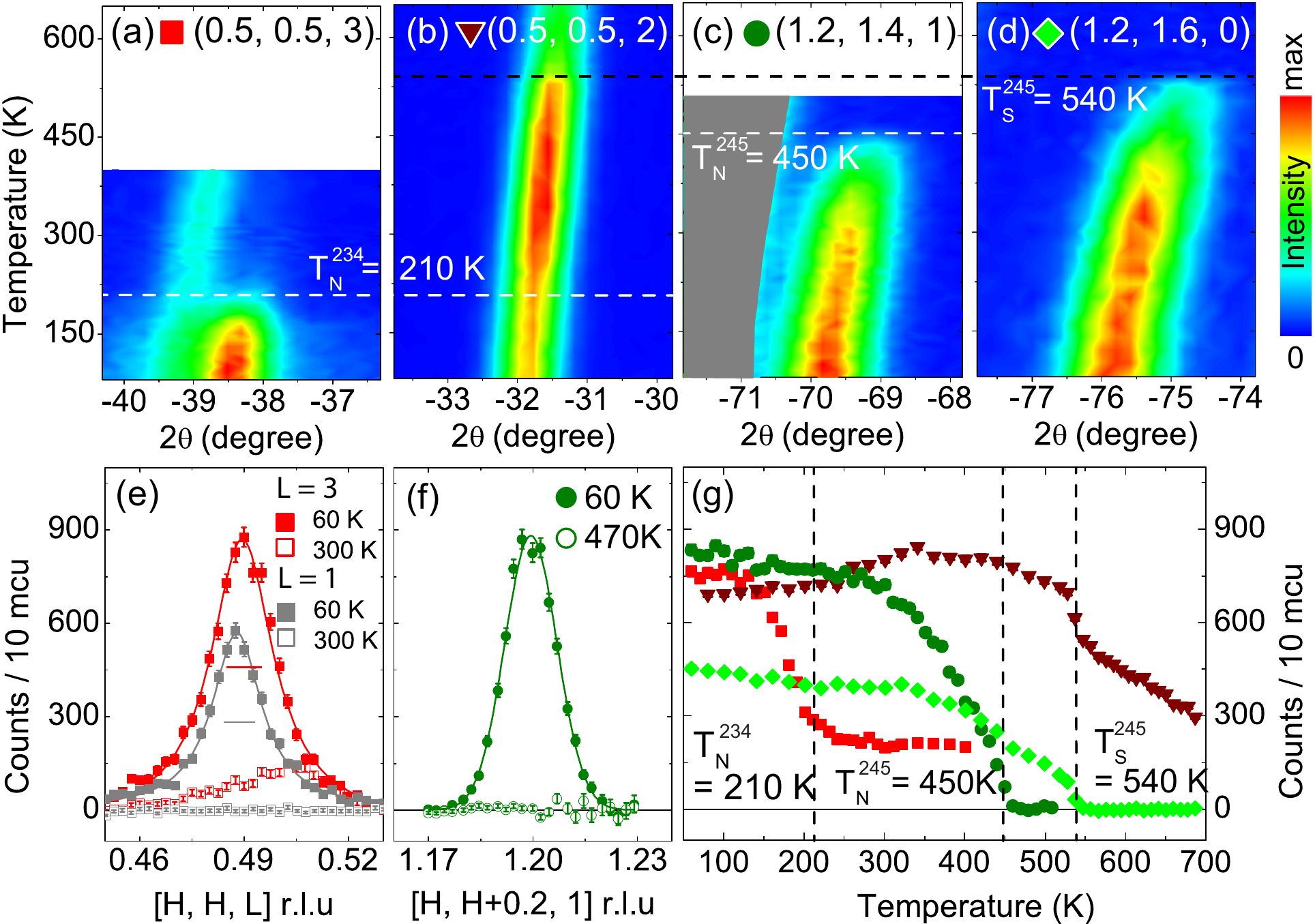}
\caption{(Color online).   (a) $\theta-2\theta$ scans of the magnetic peak at $Q=(0.5, 0.5, 3)$ and (b) the rhombic iron vacancy ordered peak at $Q = (0.5, 0.5, 2)$. The white dashed lines at $T = 210$ K in (a) and (b) mark the N$\acute{\mathrm{e}}$el temperature of the 234 phase. (c) $\theta-2\theta$ scans of the magnetic peak at $Q=(1.2, 1.4, 1)$ and (d) the $\sqrt{5} \times\sqrt{5}$ iron vacancy ordered peak at $Q=(1.2, 1.6, 0)$ of the 245 phase. The grey area in (c) blocks a spurious peak of Al. (e) Scans through the magnetic peaks associated with the 234 phase at $Q = (0.5, 0.5, L=1$ and $3)$ at 60 K and 300 K along the $[H, H, L]$ directions and (f) the magnetic peak associated with the 245 phase at $Q=(1.2, 1.4, 1)$ at 60 K and 470 K along the $[H, H+0.2, 1]$ direction, respectively. The lattice constants are chosen as the average of the two phases with a = b = 3.846 \AA\, and c = 14.188 \AA\ in the tetragonal notation. (g) Order parameters of the magnetic (red squares) and rhombic iron vacancy order (brown triangles) of the 234 phase, and the $\sqrt{5} \times\sqrt{5}$ iron vacancy order (light green diamonds) and the block AF order (dark green circles) of the 245 phase, respectively.  The intensities in (g) are obtained by fitting the heights of the peaks in (a)-(d).   }
\label{fig2}
\end{figure*}

\begin{figure*}[t]
\includegraphics[scale=0.65]{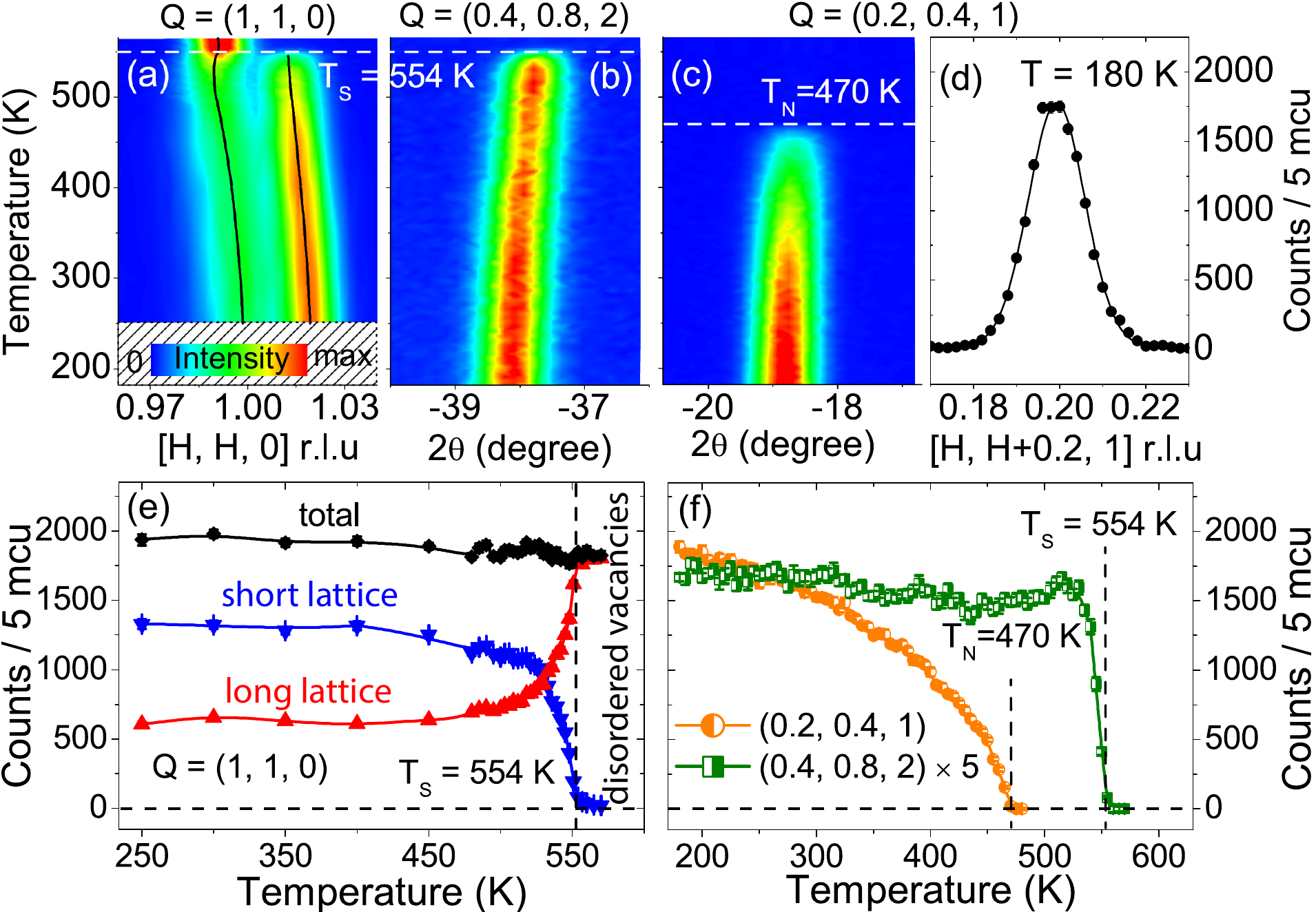}
\caption{(Color online). (a) Two well-separated peaks at $Q = (1, 1, 0)$ along the $[H, H, 0]$ direction. The left peak is associated with the 245 phase that has the lattice constants of $a = b = 3.782$ \AA\ at 250 K, and the iron vacancy disordered phase with $a = b = 3.811$ \AA\ at 570 K. The right peak belongs to the metallic phase with shorter lattice constants of $a = b = 3.705$ \AA\ at 250 K. (b) $\theta-2\theta$ scans through the $\sqrt{5} \times\sqrt{5}$ iron vacancy ordered peak at $Q = (0.4, 0.8, 2)$ and (c) the block AF ordered peak at $Q = (0.2, 0.4, 1)$ of the 245 phase. (d) A magnetic peak scan through $Q = (0.2, 0.4, 1)$ along the [H, H+0.2, 1] direction at 180 K. (e) The blue and red triangles are peak intensities extracted from the right and left peaks in (a), respectively. The black diamond is a sum of the two at each temperature. At temperatures above $T_s = 554$ K, the two phases merge into an iron vacancy disordered phase. (f) Temperature dependence of the peak heights obtained from (b) and (c), respectively, which shows the N$\acute{\mathrm{e}}$el temperature at $T_N = 470$ K and the iron vacancy order-disorder transition temperature at $T_s = 554$ K. The intensities of the peak at $Q = (0.4, 0.8, 2)$ have been multiplied by a factor of 5 in (b) and (f) for comparison.   }
\label{fig3}
\end{figure*}

We first present neutron diffraction data in Fig.~\ref{fig1} for the insulating compound with nominal composition of Rb$_{0.8}$Fe$_{1.5}$Se$_2$, in which both the 234 phase and the 245 phase may exist\cite{Zhao2012,Wangm2014}. The actual composition determined by EDX is Rb$_{0.93}$Fe$_{1.38}$Se$_2$.  We plot the peak intensities as a function of temperature at wave vectors associated with the 234 phase and the 245 phase in Fig.~\ref{fig1}(a) and the corresponding color maps in Figs.~\ref{fig1}(b)-\ref{fig1}(i). The results demonstrate that the sample consists of the 234 phase with a N$\acute{\mathrm{e}}$el temperature of $T_N$ = 220 K [Figs.~\ref{fig1}(a)-\ref{fig1}(e)] and the 245 phase with simultaneous magnetic and iron vacancy order-disorder transitions at $T_{s,N} = 400$ K [Figs.~\ref{fig1}(h)  and \ref{fig1}(i)]. The rhombic iron vacancy order of the 234 phase exists at all the temperatures we measured [Fig.~\ref{fig1}(g)]. In addition, we find another set of peaks at $H = 1/8, 3/8, 5/8$, and $7/8$ in the $[H, H, L]$ plane, where $L = odd$. The peak heights at $H=3/8, 5/8$, and $7/8$ in the $[H, H, 1]$ scattering plane follow the trend of $\mathrm{\mid\bf{S}_{\perp}\mid}^2\mathrm{F}(\bf{Q}) ^2$, where $\mathrm{\bf{S}_{\perp}}$ is the perpendicular component of in-plane ordered moments to momentum transfer $\mathrm{\bf{Q}}$ and $\mathrm{F(\bf{Q})}$ is the magnetic form factor of Fe$^{2+}$ demonstrating that they are magnetic. The temperature dependence of the peak intensities at $Q = (3/8, 3/8, 1)$ shows a transition at $T_N^{234} = 220$ K, suggesting that the new set of peaks has the same magnetic origin as the stripe AF order. One possible explanation is that the 234 phase exhibits a super iron vacancy order in addition to the rhombic iron vacancy order. Kinks are clearly observed in the intensities of the peaks associated with the 234 phase at the transition temperature $T_{s,N}^{245} = 400$ K of the 245 phase, demonstrating the mobility of the iron vacancies between the two phases. A transition-like behavior in the peak intensities at $Q=(0.5, 0.5, 3)$ at 400 K that is much higher than the N$\acute{\mathrm{e}}$el temperature of $T_N^{234}$ = 220 K indicates that the peak intensities have a combination contribution from the nuclear reflection of the super iron vacancy order and a $\sqrt{2} \times\sqrt{2}$ Rb vacancy order between 220 K and 400 K, and a pure $\sqrt{2} \times\sqrt{2}$ Rb vacancy order at temperatures above 400 K\cite{Wang2011b}. It is possible that the free iron ions released from the iron vacancy disordered 245 phase at temperatures above $T_{s,N}^{245}=400$ K occupy the vacant sites that form the super iron vacancy order. As a result, the peak intensities at $Q=(0.5, 0.5, 3)$ decrease at temperatures above $T_s=400$ K, and conversely, the peak at $Q=(0.5, 0.5, 2)$ that represents the rhombic iron vacancy order is enhanced as seen in Figs.~\ref{fig1}(a), \ref{fig1}(f), and \ref{fig1}(g). The weak Bragg peaks of the 245 phase presented in Fig.~\ref{fig1}(a) indicate that the 245 phase has a small volume fraction, consistent with the low iron content revealed by EDX and the existence of a super iron vacancy order in addition to the rhombic iron vacancy order. We calculate the spin correlation length $\xi$ of the 234 phase by fitting the peak in Fig.~\ref{fig1}(e) as a Gaussian function $I=I_0exp[-(H-H_0)^2/(2\sigma^2)]$. The full-width-half-maximum (FWHM) of the observed peak $\omega_{obs}=2\sqrt{2\mathrm{ln}2}\sigma=\sqrt{\omega_{res}^2+\omega_{int}^2}$, where $\omega_{res}$ is the instrumental resolution and $\omega_{int}$ is the intrinsic broadening of the spin-spin correlation to the FWHM of the reflection peak\cite{Wang2011}. After taking the instrumental resolution into account, the result of  $\xi =[2\sqrt{2}\mathrm{ln}2/\pi](a/\omega_{int})=150 \pm 5$ \AA\ suggests that the 234 phase and the 245 phase are separated on mesoscopic scales in Rb$_{0.8}$Fe$_{1.5}$Se$_2$.

Neutron diffraction data of a single crystal with nominal composition Rb$_{0.8}$Fe$_{1.5}$SeS, in which EDX reveals the actual composition of Rb$_{0.55}$Fe$_{1.53}$Se$_{1.25}$S$_{0.75}$, are presented in Fig.~\ref{fig2}.  The data confirm the existence of the 234 phase with $T_N^{234}$ = 210 K and the 245 phase with separated phase transition temperatures of $T_N^{245}$ = 450 K and $T_s^{245}$ = 540 K. The scans through the magnetic peaks of the 234 phase at $Q = (0.5, 0.5, L = 1$ and $3)$ in Fig.~\ref{fig2}(e) yield a spin correlation length of $\xi = 130 \pm 7$ \AA\ for $L = 1$, and $\xi = 115 \pm 5$ \AA\ for $L = 3$ at 60 K. The order parameters obtained from Figs.~\ref{fig2}(a)-\ref{fig2}(d) are presented in Fig.~\ref{fig2} (g). These data suggest that the volume fractions of the two phases are comparable. In contrast to the measurements in Rb$_{0.8}$Fe$_{1.5}$Se$_2$, we did not observe any reflection peaks indicative of the existence of a super iron vacancy order. The 234 and 245 phases are each found in Rb$_{0.8}$Fe$_{1.5}$Se$_2$, Rb$_{0.8}$Fe$_{1.5}$SeS, and the previously studied Rb$_{0.8}$Fe$_{1.5}$S$_2$\cite{Wangm2014}. This implies that the Se:S ratio does not affect the formation of the iron vacancy ordered phases.

To elucidate the role of the iron content, we carried out additional neutron diffraction measurements on the crystal with nominal composition of Rb$_{0.8}$Fe$_2$S$_2$. The real composition as revealed by EDX is Rb$_{0.75}$Fe$_{1.85}$S$_{2}$. The diffraction data are presented in Fig.~\ref{fig3}. Scans at $Q = (1, 1, 0)$ show two well separated peaks below 554 K [Fig.~\ref{fig3}(a)]. Measurements at the wave vectors associate with the 245 phase in Fig.~\ref{fig3}(b) and Fig.~\ref{fig3}(c) confirm the presence of the 245 phase with a N$\acute{\mathrm{e}}$el temperature of $T_N$ = 470 K and a structural transition temperature of $T_s = 554$ K. Using the set of parameters $a = b = 3.782$ \AA\, and $c = 14.029$ \AA\, the magnetic peak of the 245 phase at $Q = (0.2, 0.4, 1)$ is well centered at $H=0.2$ along the $[H, H+0.2, 1]$ direction at 180 K, as shown in Fig.~\ref{fig3}(d). Thus, we identify the left peak in Fig.~\ref{fig3}(a) centered at $Q = (1, 1, 0)$ as associated with the 245 phase. The 245 phase as a tetragonal structure should not result in an extra peak at $Q = (1, 1, 0)$\cite{Bao2011}. From the metallic behavior exhibited in transport and ARPES measurements on the same compound\cite{Yi2015}, we identify that the right peak at $Q = (1, 1, 0)$ in Fig.~\ref{fig3}(a) belongs to a metallic phase with the shorter in-plane lattice constant of $a = b = 3.705$ \AA\ at 250 K. Within the instrumental resolution, we could not distinguish the difference in the lattice constants along the $c$ axis. The phase with shorter in-plane lattice constant is attributed to one with fewer iron vacancies. The significantly shortened lattice of the metallic phase suggests that it is likely an iron vacancy-free phase. The metallic phase is approximately $67\%$ in volume as estimated from the intensities of the peaks at $Q = (1, 1, 0)$. The volume fraction ratio of the two phases and the averaged composition of  Rb$_{0.75}$Fe$_{1.85}$S$_{2}$ also implies that the 245 phase stabilizes at $y\approx1.6$ and the metallic phase is iron vacancy-free with $y\approx2$. The two phases merge with increasing temperature into a single iron vacancy disordered phase at temperatures just above $T_s= 554$ K, as shown in Figs.~\ref{fig3}(a) and \ref{fig3}(e). 

\begin{figure}[t]
\includegraphics[scale=0.4]{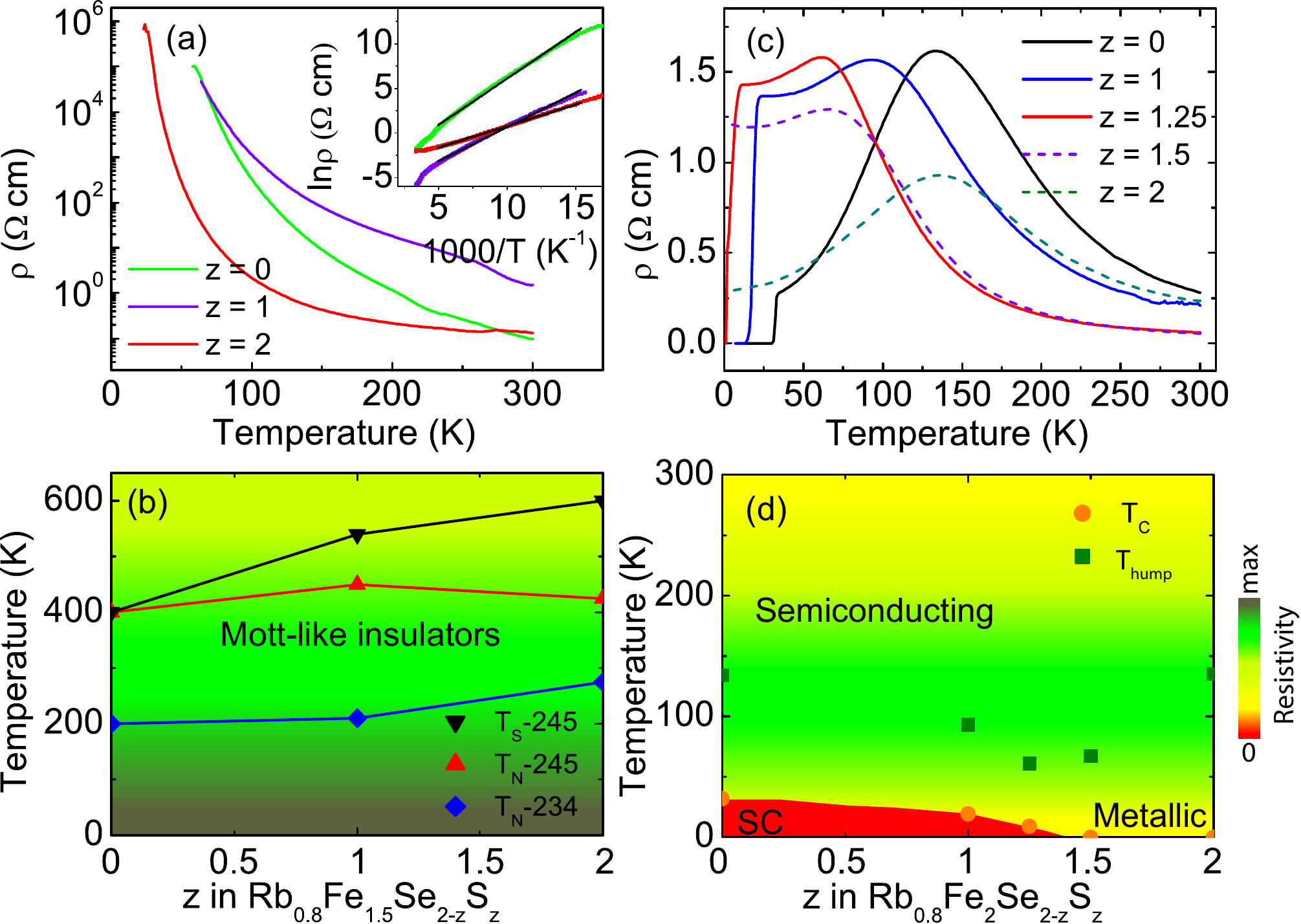}
\caption{(Color online). (a) In-plane resistivity measurements of insulating Rb$_{0.8}$Fe$_{1.5}$Se$_{2-z}$S$_z$ with $z = 0, 1$, and 2. The inset shows resistivity in log scale plotted against $1000/T$, along with fits of $\rho= \rho_0 exp(E_a/k_B T)$. (b) A phase diagram of insulating Rb$_{0.8}$Fe$_{1.5}$Se$_{2-z}$S$_z$ as a function of sulfur doping. The solid lines mark the N$\acute{\mathrm{e}}$el temperatures of the 234 phase (blue), the 245 phase (red) and the $\sqrt{5} \times\sqrt{5}$ iron vacancy order-disorder transition (black). (c) Resistivity measurements for Rb$_{0.8}$Fe$_2$Se$_{2-z}$S$_z$ with $z = 0, 1, 1.25, 1.5$ and 2. (d) A phase diagram of Rb$_{0.8}$Fe$_2$Se$_{2-z}$S$_z$ as a function of sulfur doping. The $T_c$s and $T_{hump}$s are extracted from (c). The color in (b) and (d) indicates the magnitude of resistivity. }
\label{fig4}
\end{figure}

 The results of measurements of the resistivity on samples with nominal compositions of Rb$_{0.8}$Fe$_{1.5}$Se$_{2-z}$S$_z$, z = 0, 1, 2, are presented in Fig.~\ref{fig4} (a). All of the resistivity curves turn up at low temperatures, indicating insulating behaviors. In the inset, we show the results of fits of the resistivity data to the form $\rho= \rho_0exp(E_a/k_B T)$, where $k_B$ is the Boltzmann constant and $E_a$ is the thermal activation gap\cite{Fang2010}, resulting in $E_a = 89.5 \pm 0.5, 66.1 \pm 0.4$, and $41.1 \pm 0.2$ meV for $z = 0, 1$, and 2, respectively. Consistent with previous measurements, the thermal activation gaps\cite{Zhao2012,Wangm2014,Fang2010} are much smaller than the band gaps that are measured directly by ARPES for both the 234 phase and the 245 phase\cite{Wang2015b,Chen2011a}, which could be due to the pressure effect on the internal interfaces or alternatively the existence of a small undetectable quasiparticle spectral weight near Fermi energy.  A phase diagram with the transition temperatures of the two phases obtained from neutron diffraction experiments is given in Fig.~\ref{fig4}(b).

In Fig.~\ref{fig4}(c), we show the results of resistivity measurements on various nominal compositions of Rb$_{0.8}$Fe$_2$Se$_{2-z}$S$_z$ single crystals below 300 K. The $T_c$s are gradually suppressed from 32 K ($z = 0$) to 20 K ($z = 1$) to 9 K ($z = 1.25$) and 0 K ($z = 1.5$) by sulfur substitution. Humps in the resistivity ($T_{hump}$) appear between 75 K and 150 K. The compounds are semiconducting for $T > T_{hump}$, and become superconducting or metallic for $T < T_{hump}$. The pure Rb$_{0.8}$Fe$_2$S$_2$ crystal is clearly metallic from the resistivity curve in Fig.~\ref{fig4}(c) and the band structure measurements by ARPES\cite{Yi2015}, in contrast to a previous report showing a semiconducting behavior\cite{Lei2011}; the latter could be caused by iron content deviations. The metallic characteristic in Rb$_{0.8}$Fe$_2$S$_2$ is consistent with the observation of the iron vacancy-free phase in neutron diffraction measurements, as shown in Figs.~\ref{fig3}(a) and \ref{fig3}(e). These observations result in a phase diagram as shown in Fig.~\ref{fig4}(d), suggesting that the metallic phase in Rb$_{0.8}$Fe$_2$S$_2$ is continuously connected to the SC phase in Rb$_{0.8}$Fe$_2$Se$_2$ as S is progressively replaced by Se. The commonly observed 245 phase exists in all the nominal compositions of Rb$_{0.8}$Fe$_2$Se$_{2-z}$S$_z$ as a mesoscopically separated phase. Note that the Fe content is in fact less than 2 which means that all of the samples are in the two phase coexistence region.

\section{Discussion}

\begin{figure}[t]
\includegraphics[scale=0.37]{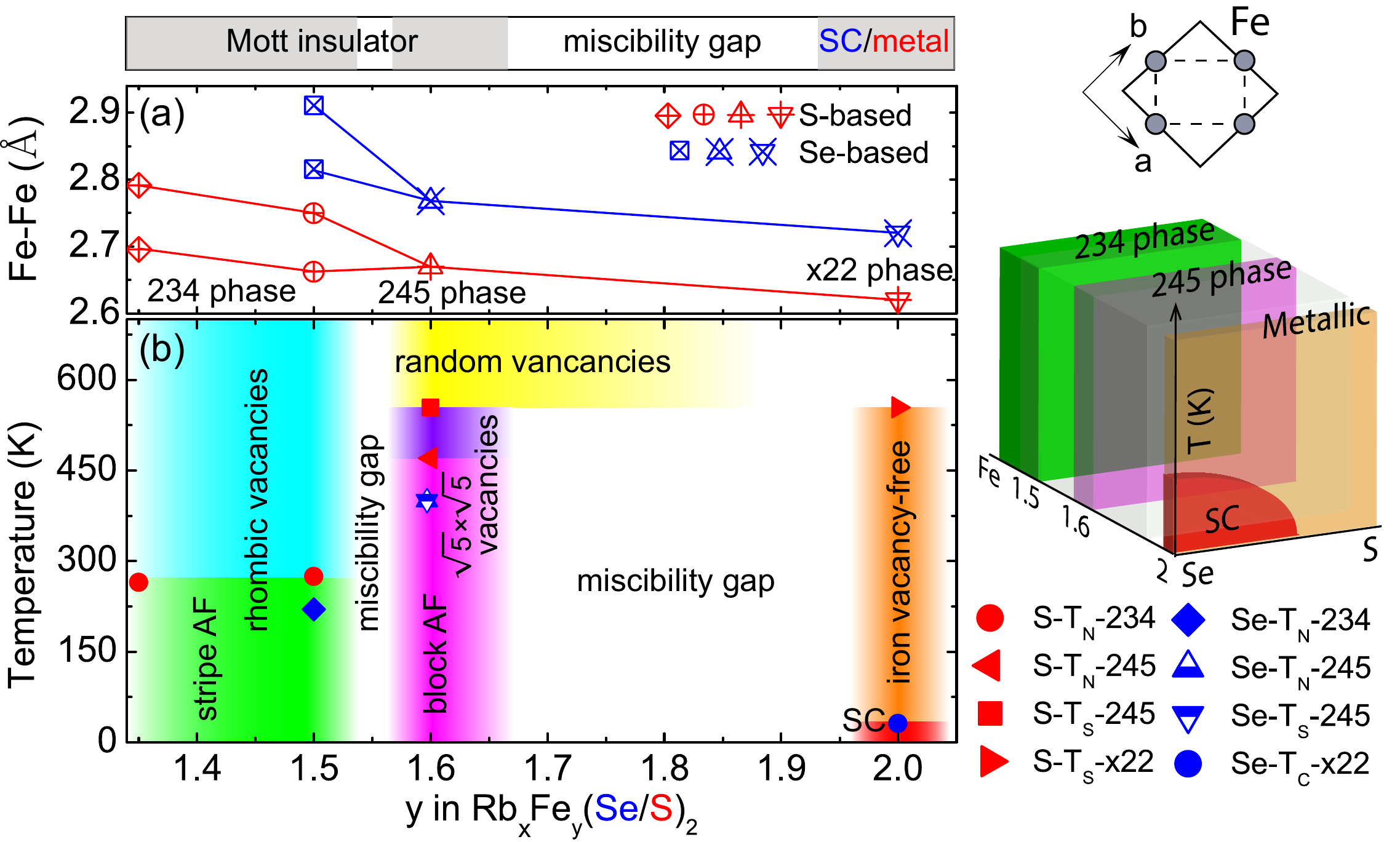}
\caption{(Color online).  (a) Fe-Fe bond lengths of the 234 phase, the 245 phase, and the ``$\mathrm{x22}$" phase for the S-based (red) and Se-based (blue) materials. The upper inset on the right is a sketch showing the Fe-Fe bonds along the two directions and the unit cell used throughout the paper marked by the solid line. (b) A phase diagram of temperature verses the iron content $y$ in Rb$_{x}$Fe$_y$(Se/S)$_2$. The red and blue points are associated with the sulfide and selenide compounds, respectively. The white areas are miscibility gaps. The point at $y=1.35$ is obtained from previous work\cite{Wang2015b}.The lower inset on the right is a schematic three-dimensional phase diagram with temperature, iron content, and Se:S ratio. }
\label{fig5}
\end{figure}

Our results demonstrate that the rhombic iron vacancy ordered phase is energetically stable for compositions with iron content $y \leq1.5$. The associated stripe AF order forms at temperatures below $T_N^{234}$. The pure stripe phase is achieved in samples with a refined composition of Rb$_{0.78}$Fe$_{1.35}$S$_2$\cite{Wang2015b}. In the regime $1.5 < y < 1.6$, both the rhombic iron vacancy order and the $\sqrt{5} \times\sqrt{5}$ iron vacancy order coexist mesoscopically, while only the ratio of the volume fraction of the two phases varies\cite{Wangm2014}, suggesting the existence of a miscibility gap. The two phases are both characterized as Mott insulators with large gaps at the Fermi level and localized electrons, promoted by the iron vacancies, as well as a high spin configuration, S = 2\cite{Wang2015,Chen2011a,Wang2011c,Yu2013}. The first order structural transition in nominal composition of Rb$_{0.8}$Fe$_2$S$_2$ at $T_s = 554$ K, and the distinct Fe-Fe bond distances[Fig.~\ref{fig5}(a)], as well as the iron contents in the 245 phase and the metallic phase imply the existence of another miscibility gap in $y$ between the two phases at temperatures below $T_s$. Adjusting the additional iron in the starting mixtures does not extend the miscibility gap\cite{Liu2012b}. The temperature dependence of the phase transitions in Rb$_{0.8}$Fe$_2$S$_2$ are in good agreement with those of the SC phase in $A_x$Fe$_{2-\delta}$Se$_2$\cite{Ricci2011a,Carr2014}.

Neutron\cite{Carr2014,Kobayashi2015} and synchrotron x-ray \cite{Shoemaker2012} diffraction, together with NMR\cite{Texier2012} studies on the SC phase of $A_x$Fe$_{2-\delta}$Se$_2$ reveal that the SC phase is nearly iron vacancy-free ($\delta\approx0$) with $x$ spanning between 0.3 and 0.58. We refer to the metallic and superconducting iron vacancy-free phases as the ``$\mathrm{x22}$" phase. Due to the existence of the $\sqrt{5} \times\sqrt{5}$ iron vacancy ordered phase, the macroscopically averaged iron content of the two phases below $T_s$ and the iron vacancy randomly distributed phase above $T_s$ are always less than $y = 2$. The absence of superconductivity in metallic Rb$_{0.8}$Fe$_2$S$_2$ is due to the weaker electron correlation effects compared to those in SC Rb$_{0.8}$Fe$_2$Se$_2$, presumably caused by the smaller sulfur ions\cite{Yi2015} and the shorter Fe-Fe bonds, as plotted in Fig.~\ref{fig5}(a). The shorter NN Fe-Fe bonds in the ``$\mathrm{x22}$" phases compared to those in the phases of $y < 1.6$ with iron vacancies are consistent with the understanding that the ``$\mathrm{x22}$" phases are nearly iron vacancy-free. The 245 phase, which has the energetically favored $\sqrt{5} \times\sqrt{5}$ iron vacancy order for iron concentrations near 1.6, supplies iron ions to stabilize the SC (Se-based) and metallic (S-based) iron vacancy-free phases. As revealed in the three dimensional phase diagram in Fig.~\ref{fig5}, isovalent substitution of selenium on the sulfur sites induces superconductivity from a nonmagnetic metallic phase. This is in clear contrast to the effects of phosphorus doping on the arsenic sites in BaFe$_2$As$_2$, where the superconductivity emerges continuously from the antiferromagnetic parent compound\cite{Johnston2010,Stewart2011}.

The effects of disorder are always more significant in quenched samples and at interfaces. Not surprisingly, some compositions within the miscibility gaps are reported in quenched samples. A scanning tunneling microscopy study reveals a 278 phase with $y = 1.75$ in some areas of SC K$_x$Fe$_y$Se$_2$\cite{Ding2013}. The 278 phase is more likely a surface state, since it has not been observed in bulk materials studied over a wide range of compositions. Neutron diffraction studies on a NSC Rb$_x$Fe$_{2-y}$Se$_2$ compound also identify a NSC phase with a composition of Rb$_{0.92}$Fe$_{1.81}$Se$_2$, containing interdigitated 234 and 245 phase\cite{Kobayashi2015}. However, a superconducting transition has been realized in an annealed NSC sample\cite{Wang2015a}. All these observations are consistent with the existence of the miscibility gaps in the phase diagram in equilibrium bulk materials.

\section{Conclusion}

Our studies on both the Se-based and S-based Rb$_x$Fe$_{y}$Se$_{2-z}$S$_z$ compounds result in phase diagrams in Fig.~\ref{fig5}, demonstrating that the iron content is a crucial parameter in stabilizing the various iron vacancy orders and the associated magnetic orders. The ratio of selenium and sulfur, on the other hand, does not affect the formation of these phases, but is important for the emergence of superconductivity. The stripe AF order, the block AF order, and the SC or metallic phase in the Rb$_x$Fe$_y$Se$_{2-z}$S$_z$ system have distinct lattice constants and are well separated in iron content. The relationship between these phases is clearly different from that between the AF phase and the SC phase in iron pnictides, where the AF order and the superconductivity microscopically coexist\cite{Stewart2011}, as there does not appear to be a continuous path that connects these magnetic phases to the superconducting phase. The existence of the miscibility gaps is the key to understanding the relationship between the various phases and, therefore, ultimately the superconductivity in these rich and interesting materials.

\section{Acknowledgments}

This work was supported by the Director, Office of Science, Office of Basic Energy Sciences, U.S. Department of Energy, under Contract No. DE-AC02-05CH11231 and the Office of Basic Energy Sciences U.S. DOE Grant No. DE-AC03-76SF008. The research at Oak Ridge National Laboratory's High-Flux Isotope Reactor is sponsored by the Scientific User Facilities Division, Office of Basic Energy Sciences, U.S. Department of Energy.

\bibliography{mengbib}

\end{document}